# Suitability of available interatomic potentials for Sn to model its 2D allotropes

Marcin Maździarz[a)]
*Institute of Fundamental Technological Research Polish Academy of Sciences, Pawińskiego 5B, 02-106 Warsaw, Poland*

The suitability of a range of interatomic potentials for elemental tin was evaluated in order to identify an appropriate potential for modeling the stanene (2D tin) allotropes. Structural and mechanical properties of the flat (F), low-buckled (LB), high-buckled (HB), full dumbbell (FD), trigonal dumbbell (TD), honeycomb dumbbell (HD) and large honeycomb dumbbell (LHD) monolayer tin (stanene) phases, were gained by means of the density functional theory (DFT) and molecular statics (MS) calculations with ten different Tersoff, modified embedded atom method (MEAM) and machine-learning-based (ML-IAP) interatomic potentials. A systematic quantitative comparison and discussion of the results are reported.

Keywords: Tin; Stanene; 2D materials; Interatomic potentials; Force fields; DFT; Mechanical properties

## I. INTRODUCTION

The scientific boom in 2D materials began in 2004 with the fabrication of graphene, a flat allotrope of carbon[1]. Immediately, researchers became interested in other 2D materials as well, and naturally gravitated toward other elements in the carbon group of the 14th (formerly IV or IVA main) group of the periodic table, i.e. silicon (Si), germanium (Ge), tin (Sn), lead (Pb) and flerovium (Fl).

The flat, 2D form of tin (Sn, Latin: *stannum*) is called stanene and is a more analogue of the silicene (2D silicon), germanene (2D germanium) and plumbene (2D lead) than graphene (2D carbon). There is no layered form of tin analogous to graphite in nature. However, for Si, Ge, Sn, Pb, and Fl, it was possible to grow their 2D allotropes on suitable substrates by epitaxy[2,3].

Stanene is seen as a promising material for new electronic and spintronic applications due to its unique physical properties such as the quantum spin Hall (QSH) effect, topological superconductivity, giant magnetoresistance, perfect spin filter, and anomalous Seebeck effect[4], and can potentially be used in electronics, optoelectronics, spintronics, thermotics, chemistry, mechanics, and sensor nanosystems[5].

A total of approximately ten hypothetical polymorphs have been proposed in the scientific literature for 2D silicon, germanium, and tin[6–8]. If we limit ourselves to monolayers, this number decreases. Using *ab initio* calculations, it was possible here to relax and then analyze seven phases of stanene, namely flat (F), low-buckled (LB), high-buckled (HB), full dumbbell (FD), trigonal dumbbell (TD), honeycomb dumbbell (HD) and large honeycomb dumbbell (LHD). Complete structural and especially mechanical data for different stanene allotropes are very limited, see[9–14].

If we could then all simulations in material modeling would be *ab initio* for sure, unfortunately their cost does not allow it and we use simplified methods like molecular dynamics/statics. A critical point in such calculations is the quality/universality/transferability of the interatomic potentials, i.e. how they behave outside the training set[15–17].

Most interatomic potentials, both classical and machine learning-based (ML-IAP), are parameterized for 3D structures. The question naturally arises whether they are suitable

[a)]Electronic mail: mmazdz@ippt.pan.pl.

for modeling their 2D allotropes. In the present research study, using *ab initio* calculations, we determined the structural and mechanical properties of seven 2D phases of Sn and then investigated whether the available potentials are able to reproduce these properties.

As far as I know, there is no publication that examines the ability of interatomic potentials for tin to describe its two-dimensional allotropes. Furthermore, there is no publication that comprehensively studies the structural and mechanical properties of stanene allotropes using DFT calculations. The objective of this paper is twofold: firstly, to determine the structural and mechanical properties of two-dimensional tin allotropes using *ab initio* methods, and secondly, to examine the ability of available interatomic potentials to reproduce these properties.

## II. METHODOLOGY

The initial stage of the study involved the generation, based on the available literature[6–8], unit cells of hypothetical allotropes of monolayer tin. Seven such phases were successfully obtained. Then, using *ab initio* computations outlined in the Section II A, structural and mechanical data for these phases were determined and analyzed. These data contain lattice parameters, average cohesion energy, average bond length, average height, and 2D elastic constants. The data thus determined were then taken as reference results and marked as $\text{Value}^{\text{DFT}}$. The same data were then computed by classical molecular statics (MS), as outlined in the Section II A, and the ten interatomic potentials enumerated in Subsection II B. These results were then marked as $\text{Value}^{\text{potential}}$. Having reference data and those from the MS calculations, we specify the mean absolute percentage error (MAPE):

$$\text{MAPE} = \frac{100\%}{n}\sum_{i=1}^{n}\left|\frac{\text{Value}^{\text{DFT}} - \text{Value}^{\text{potential}}}{\text{Value}^{\text{DFT}}}\right|, \quad (1)$$

which allows us to quantify the analyzed interatomic potentials.

Furthermore, a series of classical molecular dynamics (MD) simulations (200 atoms and 10000 time steps, NVE-microcanonical ensemble) and the built-in LAMMPS function *timesteps/s* were employed to assess the computational cost of the examined interatomic potentials. The results were then normalized against the longest simulation time. For the purposes of performance testing, a serial version of LAMMPS and a laptop equipped with an Intel Core i5-8265U processor and Linux were used. LAMMPS (Stable Release from 2 August 2023) was compiled with the GCC compiler version 11.4.

### A. Ab Initio and Atomistic Computations

The *ab initio* (relaxation of the structures, computation of cohesive energies and 2D elastic constants) and atomistic (relaxation of the structures, computation of cohesive energies and 2D elastic constants, CPU cost of the analyzed potentials) methodology here is analogous to that in[7]. The programs used are: for density functional theory (DFT)[18,19] and density functional perturbation theory (DFPT)[20] ABINIT[21] with generalized gradient approximation (GGA), PBEsol[22] as exchange-correlation (XC) functional and optimized norm-conserving Vanderbilt pseudopotential[23] (ONCVPP) taken from PseudoDojo project[24]. For molecular statics (MS) and molecular dynamics (MD), the Large-scale Atomic/Molecular Massively Parallel Simulator (LAMMPS)[25] and for visualization and analysis, the Open Visualization Tool (OVITO)[26].

### B. Interatomic Potentials

The parametrizations of the potentials enumerated hereafter were obtained from the NIST Interatomic Potentials Repository[27] and/or from LAMMPS code sources and/or through the assistance of the authors of the aforementioned publications.

1. Tersoff2016[28]: the Tersoff potential parameters fitted to an *ab-initio* derived training set that included the equation of state, cohesive energy, lattice constants, buckling height, in-plane stiffness and phonon dispersion of low-buckled stanene, calculated using density functional theory (DFT)
2. MEAM1997[29]: the modified embedded atom method (MEAM) potential fitted to cohesive energy, lattice and elastic constant of $\alpha$ and $\beta$ tin
3. MEAM2017[30]: the modified embedded atom method (MEAM) potential fitted to both solid and liquid properties of tin
4. MEAM2018a[31]: the modified embedded atom method (MEAM) potential fitted to high temperature elastic constants of tin
5. MEAM2018b[32]: the modified embedded atom method (MEAM) potential's parameters were optimized based on the force-matching method using the density functional theory (DFT) database of energies and forces of atomic configurations under different conditions
6. RANN[33]: combination of the embedded atom method (EAM) and rapid artificial neural network potential fitted to the *ab-initio* derived energies and forces of the four different solid phases of Sn
7. DP-PBE[34]: the machine-learning-based (ML-DP-PBE) potential (DeePMD[35]) obtained by training on the *ab initio* data from density functional theory calculations using the PBE[36] exchange–correlation functional
8. DP-SCAN[34]: the machine-learning-based (ML-DP-SCAN) potential (DeePMD[35]) obtained by training on the *ab initio* data from density functional theory calculations using the SCAN[37] exchange–correlation functional
9. POLY[38]: the polynomial machine-learning potential for elemental and alloy systems obtained by training on a datasets generated from density functional theory calculations for prototype structures
10. MTP[39]: the moment tensor machine-learning interatomic potential (MTP)[40] obtained by training on the *ab initio* data from density functional theory calculations for reference configurations

## III. RESULTS

Following the generation of 2D-Sn unit cells, the subsequent step was to relax and optimize these structures through *ab initio* calculations. The resulting unit cells for the seven stanene allotropes, i.e, the flat (F):(*hP2*, P6/mmm, No.191), low-buckled (LB):(*hP2*, P$\bar{3}$m1, No.164), high-buckled (HB):(*hP2*, P$\bar{3}$m1, No.164), full dumbbell (FD):(*hP3*, P$\bar{6}$m2, No.187), trigonal dumbbell (TD):(*hP7*, P$\bar{6}$2m, No.189), honeycomb dumbbell (HD):(*hP8*, P$\bar{6}$2m, No.189) and large honeycomb dumbbell (LHD):(*hP10*, P6/mmm, No.191) are depicted in Figs. 1a)-g) and the additional crystallographic data for them are stored in the Crystallographic Information Files (CIFs) in the Supplementary material V. All the stanene polymorphs analyzed here, as well as silicene and germanene[7,8], have 2D hexagonal symmetry, which should imply planar isotropy of physical properties.

The structural and mechanical properties of the seven stanene polymorphs, obtained as a result of *ab initio* calculations, are presented in Table I. These include lattice parameters, average cohesive energy, average bond length, average height, 2D elastic constants, 2D Young's modulus, Poisson's ratio, and 2D Kelvin moduli. It can be observed that the data available from other calculations for low-buckled and large honeycomb dumbbell stanene polymorphs are in reasonable agreement with the current findings. As can be also observed, all calculated 2D Kelvin moduli for all seven phases are positive, indicating the mechanical stability of all the structures[41].

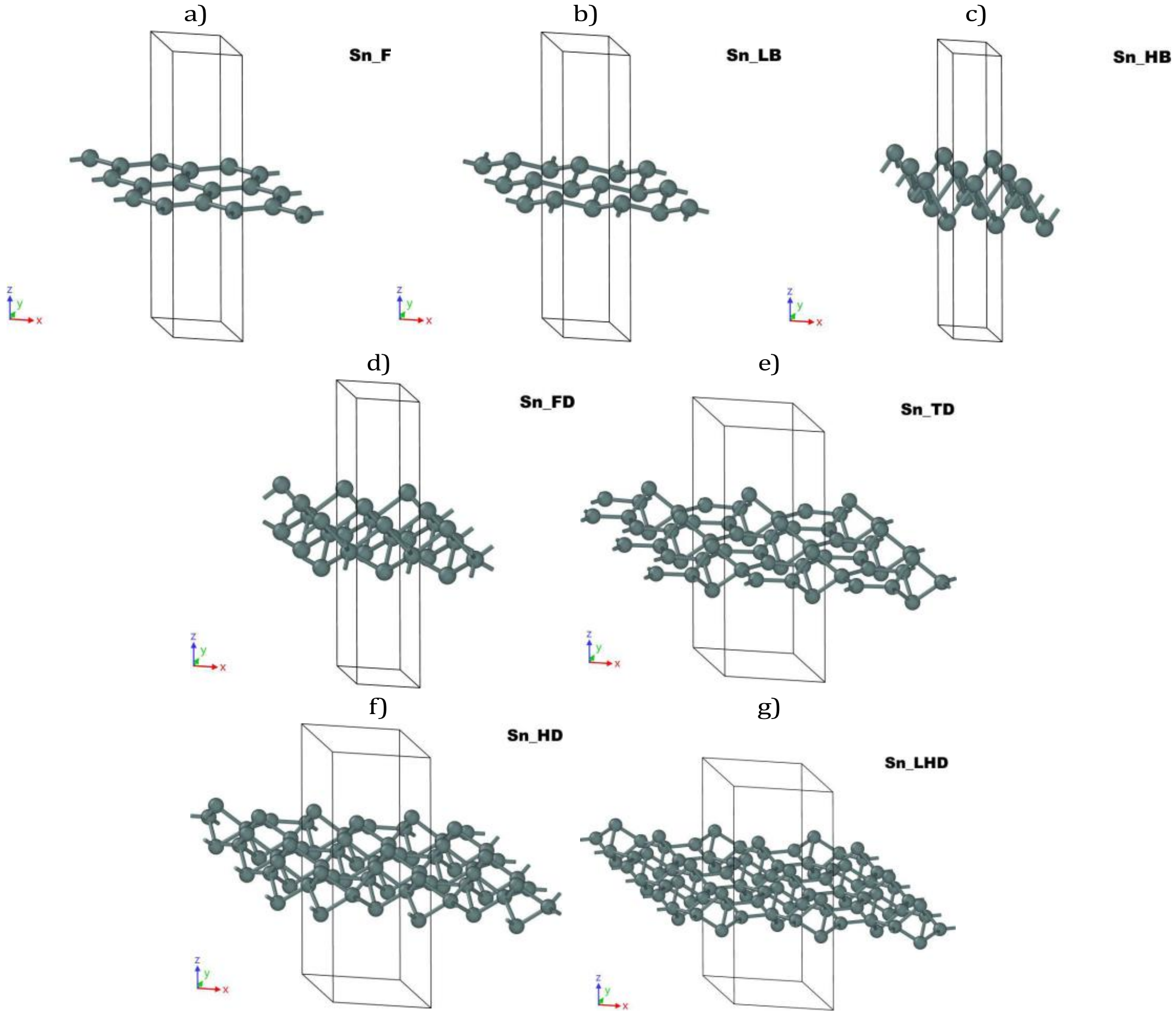

FIG. 1. Polymorphs of stanene: a) flat (F), b) low-buckled (LB), c) high-buckled (HB), d) full dumbbell (FD), e) trigonal dumbbell (TD), f) honeycomb dumbbell (HD), g) large honeycomb dumbbell (LHD).

The calculated with the use of molecular statics and the ten examined interatomic potentials for stanene (Tersoff, MEAM (×4) and machine-learning-based (ML-IAP (×5)), listed in Section II B, twelve structural and mechanical properties, i.e., lattice parameters a, b, average cohesive energy $E_c$, average bond length $d$, average height $h$, 2D elastic constants $C_{ij}$, 2D Kelvin moduli $K_i$ of the flat stanene (F) phase are gathered in Table II, of the low-buckled (LB) stanene in Table III, of the high-buckled (HB) stanene in Table IV, of full dumbbell (FD) stanene in Table V, of the trigonal dumbbell stanene (TD) in Table VI, the honeycomb dumbbell (HD) in Table VII and of the large honeycomb dumbbell stanene (LHD) in Table VIII, respectively. The aforementioned results, for each of the seven stanene allotropes, were then compared with those from the DFT computations utilizing the mean absolute percentage error (MAPE), as defined in Eq. 1.

The cumulative score of all the potentials analyzed in terms of minimum total MAPE, see Table VIII, shows the MEAM2018b potential as the best, the MEAM2017 potential as the second, and the DP-PBE potential as the third. From the point of view of computational cost in terms of relative performance measured as normalized timesteps/second in molecular dynamics (MD) simulations, MEAM2018b wins, second is Tersoff2016 and third is MEAM2017, but the differences here are minimal. Classical potentials are up to 3 orders of magnitude faster than machine-learning-based ones. Let's also look at other aspects of the performance of the potentials. Only four potentials, namely Tersoff2016, MEAM1997, MEAM2018b and POLY correctly reproduce the symmetries of all 2D-Sn phases. All five machine-learning-based potentials fail to reproduce the required isotropy of the stiffness tensor for all stanene phases, see Table VIII, either/and exhibit a spurious lack of mechanical stability, see Table II. They appear to violate the principle of symmetry (*Neumann's Principle*)[42,43], which states that the symmetry elements of any physical property of a crystal must include the symmetry elements of the crystal's point group. Since the mathematical form of such potentials is unconstrained and does not result from physical approximations, they show difficulty in extrapolating to atomic structures/compositions that differ greatly from those in the data on which they were trained[44].

TABLE I. Structural and mechanical properties of flat (F), low-buckled (LB), high-buckled (HB), full dumbbell (FD), trigonal dumbbell (TD), honeycomb dumbbell (HD) and large honeycomb dumbbell (LHD) stanene phases from density functional theory (DFT) calculations: lattice parameters a,b ( Å ), average cohesive energy $E_c$ (eV/atom), average bond length $d$ ( Å ), average height $h$ ( Å ), 2D elastic constants $C_{ij}$ (N/m), 2D Young's modulus E (N/m), Poisson's ratio $\nu$ and 2D Kelvin moduli $K_i$ (N/m).

| Polymorph | F | | LB | | HB | | FD | | TD | | HD | | LHD | |
|---|---|---|---|---|---|---|---|---|---|---|---|---|---|---|
| Source | This work | Refs. | This work | Refs. | This work | Refs. | This work | Refs. | This work | Refs. | This work | Refs. | This work | Refs. |
| a | 4.753 | | 4.594 | 4.66[b], 4.67[c] | 3.329 | 3.413[d] | 4.309 | | 7.821 | | 7.646 | 7.74[c] | 8.890 | 9.05[c] |
| b | 4.753 | | 4.594 | 4.66[b], 4.67[c] | 3.329 | 3.413[d] | 4.309 | | 7.821 | | 7.646 | 7.74[c] | 8.890 | 9.05[c] |
| $-E_c$ | 2.719 | | 2.844 | 2.68[c] | 3.179 | | 3.043 | | 2.933 | | 3.040 | 2.84[c] | 3.062 | 2.86[c] |
| $d$ | 2.746† | | 2.791 | 2.83[a], 2.83[b] | 3.266 | | 3.012 | | 2.830 | | 2.945 | | 2.897 | |
| $h$ | 0.0 | | 0.875 | 0.85[a], 0.9[b] | 2.497 | | 3.267 | | 3.514 | | 3.290 | 3.34[c] | 3.404 | 3.42[c] |
| $C_{11}$ | 31.335 | | 23.426 | 28.6[a] | 44.046 | | 15.251 | | 27.650 | | 27.923 | | 33.028 | |
| $C_{22}$ | 31.335 | | 23.426 | 28.6[a] | 44.046 | | 15.251 | | 27.650 | | 27.923 | | 33.028 | |
| $C_{12}$ | 17.704 | | 8.082 | 11.3[a] | 28.644 | | 8.710 | | 11.031 | | 13.078 | | 15.753 | |
| $C_{33}$ | 6.815 | | 7.672 | 8.65[a] | 7.701 | | 3.270 | | 8.310 | | 7.422 | | 8.637 | |
| E | 21.332 | | 20.638 | 24.14[a], 24.46[b] | 25.417 | | 10.276 | | 23.250 | | 21.797 | | 25.514 | |
| $\nu$ | 0.565 | | 0.345 | 0.395[a], 0.390[b] | 0.650 | | 0.571 | | 0.399 | | 0.468 | | 0.477 | |
| $K_I$ | 49.039 | | 31.508 | | 72.690 | | 23.961 | | 38.681 | | 41.001 | | 48.782 | |
| $K_{II}$ | 13.630 | | 15.345 | | 15.401 | | 6.540 | | 16.620 | | 14.845 | | 17.275 | |
| $K_{III}$ | 13.630 | | 15.345 | | 15.401 | | 6.540 | | 16.620 | | 14.845 | | 17.275 | |

[a] Ref. [14], [b] Ref. [13], [c] Ref. [6] [d] Ref. [10].

† An average bond lengths calculated using radial pair distribution function with a *cut-off* radius = 3.5 Å and a number of histogram bins = 1000[26].

TABLE II. Structural and mechanical properties of flat (F) stanene from molecular calculations: lattice parameters a, b ( Å ), average cohesive energy $E_c$ (eV/atom), average bond length $d$ ( Å ), average height $h$ ( Å ), 2D elastic constants $C_{ij}$ (N/m), 2D Kelvin moduli $K_i$ (N/m), mean absolute percentage error (MAPE) (%).

| Method | DFT | Tersoff 2016 | MEAM 1997 | MEAM 2017 | MEAM 2018a | MEAM 2018b | RANN 2023 | DP-PBE 2023 | DP-SCAN 2023 | POLY 2023 | MTP 2024 |
|---|---|---|---|---|---|---|---|---|---|---|---|
| a | 4.753 | 4.987 | 5.274 | 4.856 | 5.419 | 4.939 | 5.412† | 4.913 | 4.739 | 4.875 | 4.705 |
| b | 4.753 | 4.987 | 5.274 | 4.856 | 5.419 | 4.939 | 5.420† | 4.913 | 4.739 | 4.875 | 4.705 |
| $-E_C$ | 2.719 | 3.191 | 2.249 | 2.489 | 2.119 | 2.656 | 2.548 | 3.075 | 2.235 | 2.614 | 2.553 |
| $d$ | 2.746 | 2.879 | 3.047 | 2.805 | 3.092 | 2.851 | 3.126 | 2.837 | 2.735 | 2.816 | 2.718 |
| $h$ | 0.0 | 0.0 | 0.0 | 0.0 | 0.0 | 0.0 | 0.0 | 0.0 | 0.0 | 0.0 | 0.0 |
| $C_{11}$ | 31.335 | 40.547 | 25.890 | 28.729 | 24.984 | 27.073 | 9.707 | 11.347 | 165.106 | 32.077 | 49.430 |
| $C_{22}$ | 31.335 | 40.547 | 25.890 | 28.729 | 24.984 | 27.073 | 8.723 | 11.347 | 165.106 | 32.077 | 49.430 |
| $C_{12}$ | 17.704 | 16.209 | 17.699 | 10.960 | 16.023 | 19.754 | 10.765 | 13.528 | 180.506 | 30.118 | 23.780 |
| $C_{33}$ | 6.815 | 12.169 | 4.095 | 8.884 | 4.481 | 3.660 | -0.740 | -1.091 | -7.700 | 0.980 | 12.825 |
| $K_I$ | 49.039 | 56.756 | 43.590 | 39.689 | 41.007 | 46.827 | 19.969 | 24.876 | 345.611 | 62.196 | 73.211 |
| $K_{II}$ | 13.630 | 24.338 | 8.191 | 17.769 | 8.962 | 7.320 | -1.539 * | -2.181 * | -15.400 * | 1.959 | 25.650 |
| $K_{III}$ | 13.630 | 24.338 | 8.191 | 17.769 | 8.962 | 7.320 | -1.480 * | -2.181 * | -15.400 * | 1.959 | 25.650 |
| $MAPE_F$ | | 31.881 | 19.616 | 16.347 | 21.081 | 17.831 | 56.425 | 51.960 | 275.991 | 33.644 | 42.981 |

† Potential does not reproduce the correct symmetry of the structure (a/=b),
* Negative Kelvin moduli $K_i$ indicating a lack of mechanical stability.

TABLE III. Structural and mechanical properties of low-buckled (LB) stanene from molecular calculations: lattice parameters a, b ( Å ), average cohesive energy $E_c$ (eV/atom), average bond length $d$ ( Å ), average height $h$ ( Å ), 2D elastic constants $C_{ij}$ (N/m), 2D Kelvin moduli $K_i$ (N/m), mean absolute percentage error (MAPE) (%).

| Method | DFT | Tersoff 2016 | MEAM 1997 | MEAM 2017 | MEAM 2018a | MEAM 2018b | RANN 2023 | DP-PBE 2023 | DP-SCAN 2023 | POLY 2023 | MTP 2024 |
|---|---|---|---|---|---|---|---|---|---|---|---|
| a | 4.594 | 4.676 | 4.959 | 4.856† | 5.419† | 4.500 | 3.288‡ | 3.358 | 4.431 | 4.731 | 4.304 |
| b | 4.594 | 4.676 | 4.959 | 4.856† | 5.419† | 4.500 | 3.288‡ | 3.358 | 4.431 | 4.731 | 4.304 |
| $-E_C$ | 2.844 | 3.311 | 2.293 | 2.489 | 2.119 | 2.743 | 2.900 | 3.613 | 2.492 | 2.651 | 2.760 |
| $d$ | 2.791 | 2.875 | 3.019 | 2.805 | 3.092 | 2.823 | 3.326 | 3.322 | 2.924 | 2.851 | 2.777 |
| $h$ | 0.875 | 0.992 | 0.956 | 0.000 | 0.000 | 1.104 | 2.831 | 2.612 | 1.415 | 0.814 | 1.240 |
| $C_{11}$ | 23.426 | 12.678 | 12.978 | 28.729 | 24.984 | 11.730 | 48.115 | 23.558 | 90.351 | 22.791 | 16.961 |
| $C_{22}$ | 23.426 | 12.678 | 12.978 | 28.729 | 24.984 | 11.730 | 48.115 | 23.558 | 90.351 | 22.791 | 16.961 |
| $C_{12}$ | 8.082 | 3.134 | 2.502 | 10.960 | 16.023 | 3.110 | 28.561 | 18.093 | 38.954 | 16.590 | 2.235 |
| $C_{33}$ | 7.672 | 4.772 | 5.238 | 8.884 | 4.481 | 4.310 | 9.777 | 2.732 | 25.699 | 3.100 | 7.363 |
| $K_I$ | 31.508 | 15.812 | 15.480 | 39.689 | 41.007 | 14.840 | 76.675 | 41.651 | 129.305 | 39.381 | 19.196 |
| $K_{II}$ | 15.345 | 9.544 | 10.476 | 17.769 | 8.962 | 8.619 | 19.554 | 5.465 | 51.398 | 6.201 | 14.725 |
| $K_{III}$ | 15.345 | 9.544 | 10.476 | 17.769 | 8.962 | 8.619 | 19.554 | 5.465 | 51.398 | 6.201 | 14.725 |
| $MAPE_{LB}$ | | 29.385 | 29.752 | 23.219 | 36.560 | 31.725 | 82.619 | 54.066 | 171.221 | 28.022 | 19.708 |

† Input low-buckled (LB) structure converges to flat (F) one,
‡ Input low-buckled (LB) structure converges to high-buckled (HB) one.

TABLE IV. Structural and mechanical properties of high-buckled (HB) stanene from molecular calculations: lattice parameters a, b ( Å ), average cohesive energy $E_c$ (eV/atom), average bond length $d$ ( Å ), average height $h$ ( Å ), 2D elastic constants $C_{ij}$ (N/m), 2D Kelvin moduli $K_i$ (N/m), mean absolute percentage error (MAPE) (%).

| Method | DFT | Tersoff 2016 | MEAM 1997 | MEAM 2017 | MEAM 2018a | MEAM 2018b | RANN 2023 | DP-PBE 2023 | DP-SCAN 2023 | POLY 2023 | MTP 2024 |
|---|---|---|---|---|---|---|---|---|---|---|---|
| a | 3.329 | 3.043 | 3.428 | 3.300 | 3.480 | 3.232 | 3.288 | 3.358 | 3.317† | 3.297 | 3.151 |
| b | 3.329 | 3.043 | 3.428 | 3.300 | 3.480 | 3.232 | 3.288 | 3.358 | 3.332† | 3.297 | 3.151 |
| $-E_C$ | 3.179 | 2.978 | 2.582 | 2.602 | 2.343 | 2.885 | 2.900 | 3.613 | 3.180 | 2.771 | 3.477 |
| $d$ | 3.266 | 3.065 | 3.428 | 3.299 | 3.481 | 3.217 | 3.326 | 3.322 | 3.269 | 3.295 | 3.057 |
| $h$ | 2.497 | 2.565 | 2.994 | 3.092 | 3.411 | 2.585 | 2.831 | 2.612 | 2.538 | 3.442 | 2.257 |
| $C_{11}$ | 44.046 | 69.924 | 54.584 | 50.187 | 60.558 | 89.418 | 48.115 | 23.557 | 471.698 | 54.196 | 155.850 |
| $C_{22}$ | 44.046 | 69.924 | 54.584 | 50.187 | 60.558 | 89.418 | 48.115 | 23.557 | 387.154 | 54.196 | 155.850 |
| $C_{12}$ | 28.644 | 27.100 | 40.962 | 49.800 | 44.448 | 28.848 | 28.561 | 18.094 | 255.274 | 32.780 | 94.697 |
| $C_{33}$ | 7.701 | 21.412 | 6.811 | 0.194 | 8.055 | 30.285 | 9.777 | 2.731 | 75.268 | 10.708 | 30.577 |
| $K_I$ | 72.690 | 97.024 | 95.546 | 99.987 | 105.006 | 118.265 | 76.675 | 41.651 | 681.175 | 86.976 | 250.547 |
| $K_{II}$ | 15.401 | 42.824 | 13.622 | 0.388 | 16.110 | 60.570 | 19.554 | 5.463 | 177.676 | 21.415 | 61.154 |
| $K_{III}$ | 15.401 | 42.824 | 13.622 | 0.388 | 16.110 | 60.570 | 19.554 | 5.463 | 150.537 | 21.415 | 61.154 |
| $MAPE_{HB}$ | | 60.250 | 17.216 | 39.709 | 22.247 | 97.449 | 10.965 | 32.323 | 515.742 | 20.904 | 159.182 |

† Potential does not reproduce the correct symmetry of the structure (a/=b).

TABLE V. Structural and mechanical properties of full dumbbell (FD) stanene from molecular calculations: lattice parameters a, b ( Å ), average cohesive energy $E_c$ (eV/atom), average bond length $d$ ( Å ), average height $h$ ( Å ), 2D elastic constants $C_{ij}$ (N/m), 2D Kelvin moduli $K_i$ (N/m), mean absolute percentage error (MAPE) (%).

| Method | DFT | Tersoff 2016 | MEAM 1997 | MEAM 2017 | MEAM 2018a | MEAM 2018b | RANN 2023 | DP-PBE 2023 | DP-SCAN 2023 | POLY 2023 | MTP 2024 |
|---|---|---|---|---|---|---|---|---|---|---|---|
| a | 4.309 | 3.056 | 3.427 | 3.306 | 3.459 | 4.359 | 3.299 | 3.343 | 3.358 | 3.229 | 4.617 |
| b | 4.309 | 3.056 | 3.427 | 3.306 | 3.459 | 4.359 | 3.299 | 3.343 | 3.358 | 3.229 | 4.617 |
| $-E_c$ | 3.043 | 3.147 | 2.741 | 2.746 | 2.516 | 2.782 | 2.955 | 3.649 | 3.200 | 2.968 | 3.283 |
| $d$ | 3.012 | 3.074 | 3.428 | 3.306 | 3.361 | 2.998 | 3.348 | 3.355 | 3.282 | 3.303 | 2.909 |
| $h$ | 3.267 | 5.096 | 5.841 | 6.017 | 6.461 | 3.170 | 5.699 | 5.527 | 5.042 | 5.755 | 2.623 |
| $C_{11}$ | 15.251 | 98.990 | 74.550 | 70.174 | 83.734 | 24.382 | 63.948 | 35.552 | 83.271 | 109.906‡ | 34.089 |
| $C_{22}$ | 15.251 | 98.990 | 74.550 | 70.174 | 83.734 | 24.382 | 63.948 | 35.552 | 83.271 | 109.906‡ | 34.089 |
| $C_{12}$ | 8.710 | 37.917 | 56.256 | 69.053 | 60.445 | 12.875 | 42.490 | 21.910 | 53.525 | 45.523‡ | 6.230 |
| $C_{33}$ | 3.270 | 30.536 | 9.147 | 0.561 | 11.645 | 5.754 | 10.729 | 6.821 | 14.873 | 35.318‡ | 13.929 |
| $K_I$ | 23.961 | 136.907 | 130.807 | 139.227 | 144.179 | 37.257 | 106.439 | 57.462 | 136.797 | 155.429 | 40.319 |
| $K_{II}$ | 6.540 | 61.073 | 18.294 | 1.121 | 23.289 | 11.507 | 21.458 | 13.642 | 29.746 | 64.384 | 27.858 |
| $K_{III}$ | 6.540 | 61.073 | 18.294 | 1.121 | 23.289 | 11.507 | 21.458 | 13.642 | 29.746 | 70.637 | 27.858 |
| $MAPE_{FD}$ | | 411.450 | 222.917 | 208.446 | 266.200 | 42.292 | 199.120 | 93.519 | 277.683 | 472.326 | 124.271 |

‡ Potential does not reproduce the isotropy of the elasticity tensor ($2\cdot C_{33}$ /= $C_{11} - C_{12}$).

TABLE VI. Structural and mechanical properties of trigonal dumbbell (TD) stanene from molecular calculations: lattice parameters a, b ( Å ), average cohesive energy $E_c$ (eV/atom), average bond length $d$ ( Å ), average height $h$ ( Å ), 2D elastic constants $C_{ij}$ (N/m), 2D Kelvin moduli $K_i$ (N/m), mean absolute percentage error (MAPE) (%).

| Method | DFT | Tersoff 2016 | MEAM 1997 | MEAM 2017 | MEAM 2018a | MEAM 2018b | RANN 2023 | DP-PBE 2023 | DP-SCAN 2023 | POLY 2023 | MTP 2024 |
|---|---|---|---|---|---|---|---|---|---|---|---|
| a | 7.821 | 8.102 | 8.249 | 7.869 | 8.181 | 7.951 | 7.130† | 7.781 | 7.678† | 8.302 | 8.428 |
| b | 7.821 | 8.102 | 8.249 | 7.869 | 8.181 | 7.951 | 6.414† | 7.781 | 7.806† | 8.302 | 8.428 |
| $-E_c$ | 2.933 | 3.142 | 2.386 | 2.524 | 2.264 | 2.773 | 2.903 | 3.293 | 2.867 | 2.669 | 4.020 |
| $d$ | 2.830 | 2.981 | 3.050 | 2.904 | 3.122 | 2.885 | 3.238 | 2.910 | 2.918 | 3.032 | 1.718 |
| $h$ | 3.514 | 3.990 | 4.172 | 3.940 | 4.845 | 3.638 | 3.185 | 4.145 | 4.891 | 2.936 | 3.809 |
| $C_{11}$ | 27.650 | 23.369 | 19.461 | 28.514 | 21.261 | 17.086 | 21.545 | 14.818 | 33.650 | 26.747‡ | 32.672 |
| $C_{22}$ | 27.650 | 23.369 | 19.461 | 28.514 | 21.261 | 17.086 | 27.134 | 14.818 | 52.417 | 26.747‡ | 32.672 |
| $C_{12}$ | 11.031 | -0.219 | 8.511 | 11.929 | 9.824 | 8.357 | 8.056 | 13.058 | 24.280 | 24.511‡ | 20.993 |
| $C_{33}$ | 8.310 | 11.794 | 5.475 | 8.292 | 5.719 | 4.365 | 8.288 | 0.880 | 3.980 | 2.985‡ | 5.840 |
| $K_I$ | 38.681 | 23.588 | 27.972 | 40.442 | 31.085 | 25.443 | 31.895 | 27.876 | 65.427 | 51.258 | 53.665 |
| $K_{II}$ | 16.620 | 23.150 | 10.950 | 16.585 | 11.437 | 8.729 | 16.784 | 1.761 | 20.639 | 2.236 | 11.680 |
| $K_{III}$ | 16.620 | 23.588 | 10.950 | 16.585 | 11.437 | 8.729 | 16.575 | 1.761 | 7.960 | 5.970 | 11.680 |
| $MAPE_{TD}$ | | 27.356 | 22.347 | 4.121 | 20.876 | 24.294 | 10.130 | 36.787 | 39.625 | 35.067 | 29.567 |

† Potential does not reproduce the correct symmetry of the structure (a/=b),
‡ Potential does not reproduce the isotropy of the elasticity tensor ($2\cdot C_{33}$ /= $C_{11} - C_{12}$).

TABLE VII. Structural and mechanical properties of honeycomb dumbbell (HD) stanene from molecular calculations: lattice parameters a, b ( Å ), average cohesive energy $E_c$ (eV/atom), average bond length $d$ ( Å ), average height $h$ ( Å ), 2D elastic constants $C_{ij}$ (N/m), 2D Kelvin moduli $K_i$ (N/m), mean absolute percentage error (MAPE) (%).

| Method | DFT | Tersoff 2016 | MEAM 1997 | MEAM 2017 | MEAM 2018a | MEAM 2018b | RANN 2023 | DP-PBE 2023 | DP-SCAN 2023 | POLY 2023 | MTP 2024 |
|---|---|---|---|---|---|---|---|---|---|---|---|
| a | 7.646 | 7.683 | 7.769 | 7.524 | 7.023† | 7.730 | 8.007 | 9.741† | 8.659† | 7.435 | 7.953 |
| b | 7.646 | 7.683 | 7.769 | 7.524 | 7.021† | 7.730 | 8.007 | 4.782† | 7.380† | 7.435 | 7.953 |
| $-E_c$ | 3.040 | 2.881 | 2.401 | 2.468 | 2.292 | 2.788 | 2.845 | 3.579 | 2.894 | 2.820 | 3.315 |
| $d$ | 2.945 | 3.031 | 3.096 | 3.005 | 3.344 | 2.966 | 2.937 | 3.176 | 3.039 | 3.212 | 2.831 |
| $h$ | 3.290 | 3.797 | 3.970 | 3.885 | 5.429 | 3.288 | 2.943 | 4.505 | 3.186 | 3.903 | 2.607 |
| $C_{11}$ | 27.923 | 44.952 | 14.868 | 24.448 | 10.715 | 17.803 | 36.010 | 19.795 | 41.728 | 15.973‡ | 36.693 |
| $C_{22}$ | 27.923 | 44.952 | 14.868 | 24.448 | 10.729 | 17.803 | 36.010 | 14.854 | 2.145 | 18.338‡ | 36.693 |
| $C_{12}$ | 13.078 | 4.265 | 10.089 | 10.562 | 10.949 | 7.318 | 8.930 | 3.823 | 25.743 | 8.905‡ | 18.224 |
| $C_{33}$ | 7.422 | 20.344 | 2.390 | 6.943 | -0.113 | 5.243 | 13.540 | 2.198 | 19.459 | 2.543‡ | 9.234 |
| $K_I$ | 41.001 | 49.217 | 24.957 | 35.010 | 21.671 | 25.120 | 44.941 | 20.242 | 38.398 | 25.982 | 54.917 |
| $K_{II}$ | 14.845 | 40.687 | 4.779 | 13.886 | -0.227 | 10.485 | 27.080 | 14.407 | 5.475 | 8.329 | 18.469 |
| $K_{III}$ | 14.845 | 40.687 | 4.779 | 13.886 | -0.226 | 10.485 | 27.080 | 4.395 | 38.919 | 5.085 | 18.469 |
| $MAPE_{HD}$ | | 63.015 | 34.079 | 10.019 | 50.892 | 21.219 | 31.097 | 39.033 | 55.026 | 30.125 | 20.920 |

† Potential does not reproduce the correct symmetry of the structure (a/=b),
‡ Potential does not reproduce the isotropy of the elasticity tensor ($C_{11}$ /= $C_{22}$ and $2\cdot C_{33}$ /= $C_{11} - C_{12}$),
⋆ Negative Kelvin moduli $K_i$ indicating a lack of mechanical stability.

TABLE VIII. Structural and mechanical properties of large honeycomb dumbbell (LHD) stanene from molecular calculations: lattice parameters a, b (Å), average cohesive energy $E_c$ (eV/atom), average bond length $d$ (Å), average height $h$ (Å), 2D elastic constants $C_{ij}$ (N/m), 2D Kelvin moduli $K_i$ (N/m), mean absolute percentage error (MAPE) (%), relative performance measured as normalized timesteps/second in molecular dynamics (MD) simulation.

| Method | DFT | Tersoff 2016 | MEAM 1997 | MEAM 2017 | MEAM 2018a | MEAM 2018b | RANN 2023 | DP-PBE 2023 | DP-SCAN 2023 | POLY 2023 | MTP 2024 |
|---|---|---|---|---|---|---|---|---|---|---|---|
| a | 8.890 | 9.175 | 9.090 | 8.788 | 8.977 | 8.907 | 9.467 | 9.103 | 8.220 | 9.017 | 8.724† |
| b | 8.890 | 9.175 | 9.090 | 8.788 | 8.977 | 8.907 | 9.467 | 9.103 | 8.220 | 9.017 | 8.821† |
| $-E_c$ | 3.062 | 3.128 | 2.459 | 2.548 | 2.322 | 2.830 | 2.792 | 3.399 | 2.981 | 2.771 | 1762.220• |
| $d$ | 2.897 | 3.037 | 3.046 | 2.938 | 3.130 | 2.894 | 7.508 | 2.968 | 3.066 | 2.922 | 1.570 |
| $h$ | 3.404 | 3.976 | 4.153 | 3.968 | 4.853 | 3.617 | 2.932 | 3.492 | 5.058 | 3.638 | 2.515 |
| $C_{11}$ | 33.028 | 62.603 | 21.657 | 28.337 | 23.896 | 16.664 | 23.408‡ | 22.768 | 148.381‡ | 7.968‡ | 7739.930• |
| $C_{22}$ | 33.028 | 62.603 | 21.657 | 28.337 | 23.896 | 16.664 | 22.444‡ | 22.768 | 139.851‡ | 9.575‡ | 9143.533• |
| $C_{12}$ | 15.753 | 25.854 | 10.222 | 11.978 | 7.496 | 7.277 | 18.709‡ | 14.050 | 106.473‡ | 15.109‡ | 4278.919• |
| $C_{33}$ | 8.637 | 18.374 | 5.718 | 8.180 | 8.200 | 4.693 | 1.940‡ | 4.359 | 16.008‡ | -1.213‡ | 2009.998• |
| $K_I$ | 48.782 | 88.457 | 31.878 | 40.315 | 31.393 | 23.941 | 41.630 | 36.818 | 250.503 | 23.859 | 12662.705 |
| $K_{II}$ | 17.275 | 36.748 | 11.435 | 16.360 | 16.400 | 9.387 | 4.223 | 8.718 | 37.729 | -6.315⋆ | 4220.758 |
| $K_{III}$ | 17.275 | 36.748 | 11.435 | 16.360 | 16.400 | 9.387 | 3.879 | 8.718 | 32.016 | -2.427⋆ | 4019.996 |
| $MAPE_{LHD}$ | | 57.744 | 24.280 | 10.221 | 19.605 | 29.589 | 43.341 | 22.240 | 168.606 | 48.901 | 19337.268 |
| MAPE | | 681.081 | 370.206 | 312.082 | 437.462 | 264.400 | 433.696 | 329.928 | 1503.894 | 668.989 | 19733.899 |
| timesteps/s | | 4989.047 | 2159.590 | 4552.066 | 1980.937 | 5539.323 | 96.654 | 1.000 | 5.987 | 6.232 | 105.317 |

† Potential does not reproduce the correct symmetry of the structure (a/=b),

‡ Potential does not reproduce the isotropy of the elasticity tensor ($C_{11}$ /= $C_{22}$ and 2·$C_{33}$ /= $C_{11} - C_{12}$),

⋆ Negative Kelvin moduli $K_i$ indicating a lack of mechanical stability,

• Highly overstated value.

## IV. CONCLUSIONS

In the paper, a systematic quantitative comparison of different interatomic potentials of tin was carried out to reproduce the properties of seven allotropes of stanene (2D tin). In order to facilitate a comparative analysis of the ten interatomic potentials listed in Section II B, the structural and mechanical properties of the flat (F), low-buckled (LB), high-buckled (HB), full dumbbell (FD), trigonal dumbbell (TD), honeycomb dumbbell (HD) and large honeycomb dumbbell (LHD) (Figs. 1a)-g)) obtained by means of density functional theory (DFT) and molecular statics (MS) computations were used. Additionally, the computational cost and performance of the tested potentials were benchmarked.

- Only the Tersoff2016, MEAM1997 and MEAM2018b potentials correctly reconstruct the symmetry of crystals and their physical properties, see Tables II-VIII;
- The MEAM2018b potential ensures the best quantitative performance as measured by the total mean absolute percentage error (MAPE), see Table VIII;
- All five machine-learning-based potentials do not perform well in describing the mechanical properties of the seven polymorphs of stanane;
- The machine-learning-based interatomic potentials, when evaluated according to the methodology employed, are not demonstrably superior in performance (MAPE) to classical potentials. In fact, they are three orders of magnitude more computationally expensive, as evidenced Table VIII;
- In consideration of the performance, accuracy, and cost of computation, the classical potentials of the MEAM type appear to be the optimal choice in this context. Despite the absence of data for different polymorphs of stanene in the optimization of these potentials, they demonstrated the ability to reproduce their properties accurately. This is a consequence of their foundation in physical principles, their capacity for natural extrapolation, and their avoidance of mere interpolation.

The results obtained here confirm the previous observations[7,8,17] that the machine-learning-based interatomic potentials, in accordance with the methodology applied, are not superior to classical potentials in terms of their accuracy/performance (MAPE) and are meanwhile up to orders of magnitude more computationally expensive, see Table VIII. In general, classical, physics-based potentials demonstrate better universality/transferability than purely interpolative potentials based on machine learning.

It is my hope that the findings presented here will prove useful to other researchers in the selection of suitable potentials for their purposes, as well as in the parametrization of new potentials for monolayer structures.

## V. SUPPLEMENTARY MATERIAL

Crystallographic Information Files (CIFs) for polymorphs of stanene (created by qAgate: Open-source software to post-process ABINIT) are available online at Supplementary material.

- flat (F) stanene:

```
# Sn_F 'P6/MMM' 191 hexagonal
data_Sn
_symmetry_space_group_name_H-M    'P 1'
_cell_length_a    4.75336890
_cell_length_b    4.75336890
_cell_length_c    20.00000009
_cell_angle_alpha    90.00000000
_cell_angle_beta    90.00000000
_cell_angle_gamma    120.00000000
_symmetry_Int_Tables_number    1
_chemical_formula_structural    Sn
_chemical_formula_sum    Sn2
_cell_volume    391.34849752
_cell_formula_units_Z    2
loop_
 _symmetry_equiv_pos_site_id
 _symmetry_equiv_pos_as_xyz
  1  'x, y, z'
loop_
 _atom_site_type_symbol
 _atom_site_label
 _atom_site_symmetry_multiplicity
 _atom_site_fract_x
 _atom_site_fract_y
 _atom_site_fract_z
 _atom_site_occupancy
  Sn  Sn0  1  0.33333333  0.66666667  0.50000000  1
  Sn  Sn1  1  0.66666667  0.33333333  0.50000000  1
```

- low-buckled (LB) stanene:

```
# SN_LB 'P-3M1' 164 trigonal
data_Sn
_symmetry_space_group_name_H-M    'P 1'
_cell_length_a    4.59395779
_cell_length_b    4.59395779
_cell_length_c    20.00000009
_cell_angle_alpha    90.00000000
_cell_angle_beta    90.00000000
_cell_angle_gamma    120.00000000
_symmetry_Int_Tables_number    1
_chemical_formula_structural    Sn
_chemical_formula_sum    Sn2
_cell_volume    365.53976596
_cell_formula_units_Z    2
loop_
 _symmetry_equiv_pos_site_id
 _symmetry_equiv_pos_as_xyz
  1  'x, y, z'
```

```
loop_
  _atom_site_type_symbol
  _atom_site_label
  _atom_site_symmetry_multiplicity
  _atom_site_fract_x
  _atom_site_fract_y
  _atom_site_fract_z
  _atom_site_occupancy
   Sn   Sn0   1   0.33333333   0.66666667   0.47812944   1
   Sn   Sn1   1   0.66666667   0.33333333   0.52187056   1
```

- high-buckled (HB) stanene:

```
# Sn_HB 'P-3M1' 164 trigonal
data_Sn
_symmetry_space_group_name_H-M      'P 1'
_cell_length_a      3.32896256
_cell_length_b      3.32896256
_cell_length_c      20.00000009
_cell_angle_alpha     90.00000000
_cell_angle_beta     90.00000000
_cell_angle_gamma    120.00000000
_symmetry_Int_Tables_number     1
_chemical_formula_structural     Sn
_chemical_formula_sum      Sn2
_cell_volume     191.94572759
_cell_formula_units_Z      2
loop_
  _symmetry_equiv_pos_site_id
  _symmetry_equiv_pos_as_xyz
   1   'x, y, z'
loop_
  _atom_site_type_symbol
  _atom_site_label
  _atom_site_symmetry_multiplicity
  _atom_site_fract_x
  _atom_site_fract_y
  _atom_site_fract_z
  _atom_site_occupancy
   Sn   Sn0   1   0.33333333   0.66666667   0.56241445   1
   Sn   Sn1   1   0.66666667   0.33333333   0.43758555   1
```

- full dumbbell (FD) stanene:

```
# Sn_FD 'P-6m2' 187 hexagonal
data_Sn
_symmetry_space_group_name_H-M      'P 1'
_cell_length_a    4.30887239
_cell_length_b    4.30887239
_cell_length_c    20.00000009
_cell_angle_alpha      90.00000000
_cell_angle_beta    90.00000000
_cell_angle_gamma      120.00000000
_symmetry_Int_Tables_number     1
_chemical_formula_structural     Sn
_chemical_formula_sum      Sn3
```

```
_cell_volume     321.57915851
_cell_formula_units_Z      3
loop_
 _symmetry_equiv_pos_site_id
 _symmetry_equiv_pos_as_xyz
  1  'x, y, z'
loop_
 _atom_site_type_symbol
 _atom_site_label
 _atom_site_symmetry_multiplicity
 _atom_site_fract_x
 _atom_site_fract_y
 _atom_site_fract_z
 _atom_site_occupancy
  Sn  Sn0  1  0.33333333  0.66666667  0.41833575  1
  Sn  Sn1  1  0.66666667  0.33333333  0.50000000  1
  Sn  Sn2  1  0.33333333  0.66666667  0.58166425  1
```

- trigonal dumbbell (TD) stanene:

```
# Sn_P3xP3td 'P-62M' 189 hexagonal
data_Sn
_symmetry_space_group_name_H-M      'P 1'
_cell_length_a      7.82139920
_cell_length_b      7.82139920
_cell_length_c      20.00000009
_cell_angle_alpha      90.00000000
_cell_angle_beta      90.00000000
_cell_angle_gamma    120.00000000
_symmetry_Int_Tables_number     1
_chemical_formula_structural     Sn
_chemical_formula_sum      Sn7
_cell_volume     1059.56971067
_cell_formula_units_Z      7
loop_
 _symmetry_equiv_pos_site_id
 _symmetry_equiv_pos_as_xyz
  1  'x, y, z'
loop_
 _atom_site_type_symbol
 _atom_site_label
 _atom_site_symmetry_multiplicity
 _atom_site_fract_x
 _atom_site_fract_y
 _atom_site_fract_z
 _atom_site_occupancy
  Sn  Sn0  1  -0.00000000 0.00000000  0.50000000  1
  Sn  Sn1  1  -0.03884085 0.33333333  0.50000000  1
  Sn  Sn2  1  0.37217418  0.03884085  0.50000000  1
  Sn  Sn3  1  0.66666667  0.33333333  0.41216092  1
  Sn  Sn4  1  0.33333333  0.66666667  0.50000000  1
  Sn  Sn5  1  0.66666667  0.62782582  0.50000000  1
  Sn  Sn6  1  0.66666667  0.33333333  0.58783908  1
```

- honeycomb dumbbell (HD) stanene:

```
# Sn_P3xP3hd 'P-62M' 189 hexagonal
data_Sn
_symmetry_space_group_name_H-M    'P 1'
_cell_length_a    7.64601637
_cell_length_b    7.64601637
_cell_length_c    20.00000009
_cell_angle_alpha    90.00000000
_cell_angle_beta    90.00000000
_cell_angle_gamma    120.00021772
_symmetry_Int_Tables_number    1
_chemical_formula_structural    Sn
_chemical_formula_sum    Sn8
_cell_volume    1012.58181347
_cell_formula_units_Z    8
loop_
 _symmetry_equiv_pos_site_id
 _symmetry_equiv_pos_as_xyz
  1  'x, y, z'
loop_
 _atom_site_type_symbol
 _atom_site_label
 _atom_site_symmetry_multiplicity
 _atom_site_fract_x
 _atom_site_fract_y
 _atom_site_fract_z
 _atom_site_occupancy
  Sn  Sn0  1  -0.00000831  -0.00000831  0.50000000  1
  Sn  Sn1  1  -0.00000211  0.36326058  0.50000000  1
  Sn  Sn2  1  0.36326058  -0.00000211  0.50000000 1
  Sn  Sn3  1  0.66666178  0.33332813  0.41774698  1
  Sn  Sn4  1  0.33332813  0.66666178  0.41774698  1
  Sn  Sn5  1  0.63674001  0.63674001  0.50000000  1
  Sn  Sn6  1  0.66666178  0.33332813  0.58225302  1
  Sn  Sn7  1  0.33332813  0.66666178  0.58225302  1
```

- large honeycomb dumbbell (LHD) stanene:

```
# Sn_P3xP3lhd 'P6/MMM' 191 hexagonal
data_Sn
_symmetry_space_group_name_H-M    'P 1'
_cell_length_a    8.89033525
_cell_length_b    8.89033525
_cell_length_c    20.00000009
_cell_angle_alpha    90.00000000
_cell_angle_beta    90.00000000
_cell_angle_gamma    120.00000000
_symmetry_Int_Tables_number    1
_chemical_formula_structural    Sn
_chemical_formula_sum    Sn10
_cell_volume    1368.97937660
_cell_formula_units_Z    10
loop_
 _symmetry_equiv_pos_site_id
 _symmetry_equiv_pos_as_xyz
  1  'x, y, z'
loop_
```

```
_atom_site_type_symbol
_atom_site_label
_atom_site_symmetry_multiplicity
_atom_site_fract_x
_atom_site_fract_y
_atom_site_fract_z
_atom_site_occupancy
 Sn  Sn0  1  0.18240377  0.36480755  0.50000000  1
 Sn  Sn1  1  0.36480755  0.18240377  0.50000000  1
 Sn  Sn2  1  0.18240377  0.81759623  0.50000000  1
 Sn  Sn3  1  0.33333333  0.66666667  0.41489168  1
 Sn  Sn4  1  0.66666667  0.33333333  0.41489168  1
 Sn  Sn5  1  0.81759623  0.18240377  0.50000000  1
 Sn  Sn6  1  0.63519245  0.81759623  0.50000000  1
 Sn  Sn7  1  0.81759623  0.63519245  0.50000000  1
 Sn  Sn8  1  0.33333333  0.66666667  0.58510832  1
 Sn  Sn9  1  0.66666667  0.33333333  0.58510832  1
```

## ACKNOWLEDGMENTS

This work was supported by the National Science Centre (NCN – Poland) Research Project: No. 2021/43/B/ST8/03207. The computational assistance was granted through the computing cluster GRAFEN at Biocentrum Ochota, the Interdisciplinary Centre for Mathematical and Computational Modelling of Warsaw University (ICM UW) and Poznań Supercomputing and Networking Center (PSNC).

## AUTHOR DECLARATION

### Conflict of Interest

The author declare that they have no known competing financial interests or personal relationships that could have appeared to influence the work reported in this paper.